\documentstyle[11pt]{article}          
\begin{document}
\title{Photon versus Hadron Interferometry\thanks{Invited
talk at the CRIS98 meeting on HBT interferometry and Heavy
Ion Physics, Acicastello,
June 1998}}
\author{R. M. Weiner\thanks{E.Mail:
weiner@mailer.uni-marburg.de}
\date{Physics Department, University of Marburg,\\ Wieselacker
8, 35041 
Marburg, Germany \\and \\
Laboratoire de Physique Th\'eorique et Hautes 
\'Energies,
Univ. Paris-Sud,\\ 177 rue de Lourmel, 75015 Paris, France}}
\maketitle
\thispagestyle{empty}
\begin{center}
{\large\bf Abstract}\\[2ex]
\end{center}
\noindent

%%%%%%%%%%%%%%%%%%%%%%%%%%%%%%  Abstract here  %%%%%%%%%%%%%%%%%%%

Photons and mesons are both bosons and therefore satisfy the
same Bose-Einstein statistics. This leads to certain
similarities in the corresponding Bose-Einstein correlations
which underly photon and hadron intensity interferometry.
However there are also important differences 
between the two effects and these will be analyzed in the 
following.

\newcommand{\lraw}{\leftrightarrow}
\newcommand{\mbold}[1]{\mbox{\boldmath $ #1 $}}
\newcommand{\be}{\begin{equation}}
\newcommand{\ee}{\end{equation}}
\newcommand{\benn}{\begin{displaymath}}		% the same as 'equation' but
\newcommand{\eenn}{\end{displaymath}}		% without formula numbers
\newcommand{\ba}{\begin{eqnarray}}
\newcommand{\ea}{\end{eqnarray}}
\newcommand{\bann}{\begin{eqnarray*}}	% the same as 'eqnarray' but
\newcommand{\eann}{\end{eqnarray*}}
\let\ul\underline
\renewcommand{\theequation}{\arabic{section}.\arabic{equation}}
\renewcommand{\baselinestretch}{1.5}
%\begin{document}
%\maketitle

%\tableofcontents

%\newpage

\section{Introduction} 
Hanbury-Brown and Twiss (HBT) \cite{HBT} developed in the 
mid fifties the method of photon 
intensity interferometry to be used as an 
alternative to the amplitude
interferometry of Michelson. Initially this ``alternative"
was considered merely as a technical improvement 
of interest only for astronomy and it is therefore not
surprising that Goldhaber, Goldhaber
Lee and Pais (GGLP) \cite{GGLP} were not aware of the 
HBT experiment
when they discovered in 1959-1960 that pairs of identical 
pions were bunched and interpreted this effect as due to
Bose-Einstein correlations. This initial separation 
\footnote{As far as we can gather the link
between the two experiments is mentioned for the first time
in ref. \cite{Grishin}.} between
the two developments is in part due to the fact that the
 techniques used in the original HBT experiment and in
the GGLP experiment were very different:
 the HBT interferometry in astronomy consists in
measurements of distance correlations (actually
correlations of time arrivals) in order to determine (angular)
diameters of stars, while in GGLP experiments one measures
momentum correlations in order to derive radii and lifetimes
of sources of elementary particles. 
  
On the other hand for (some) people  working in optics it did not take much 
time to realize the quantum statistical significance of the
HBT experiment and it turned out that the apparently small step in the 
history of interferometry due to HBT 
represented a huge step in the history of physics, 
leading to the creation of quantum optics with all its
theoretical and practical developments.
The implications and the importance of the HBT and GGLP effects 
for particle physics were appreciated only much later. It is therefore very timely that
a conference like the present one where astronomers,
 particle and nuclear physicists meet, is
organized. At present the HBT/GGLP effect is an important tool
in particle and nuclear physics, being 
the only direct experimental method known so far for the determination of
space-time characterisitcs of particle sources. Moreover,
the phenomenon of 
Bose-Einstein correlations (BEC) presents interesting and
important theoretical problems in itself and it is thus
understandable that in the last decenium of this century
it has become an independent subject of research
\footnote{From 1990 meetings dedicated (almost) entirely 
to this subject were hold, beginning with CAMP \cite{CAMP}}.
  
Although both the HBT effect in quantum optics and in astronomy
use photons, quantum optics, being a microscopic discipline,
 is of course much more related to particle physics than
to astronomy. Among other things, in quantum optics, too, one
measures momenta, rather than distance correlations. 
On the other hand photon interferometry is not restricted
only to astronomy and quantum optics, but finds applications
also in particle and nuclear physics. As a matter of fact,
photon interferometry in particles physics is from a 
certain point of view superior to 
 hadron interferometry, because photons are weakly
interacting particles, while hadrons interact strongly.
This has two important consequences in photon BEC: (i) there
is (up to higher order corrections) no final state
interaction between photons, so that the BEC effect is
``clean"; (ii) in a high energy reaction, hadrons are
produced only at the end of the reaction (at
freeze-out), while photons from the beginning, so that
photons can provide unique information about the initial
state. For the search of quark-gluon plasma this is
essential, because if such a state of matter is formed, then
this happens only in the early stages of the reaction. 
This is also important in lower energy heavy ion reactions
where the dynamics of the reaction as well as its space-time 
geometry are studied in this way (cf. the talk by
R. Barbera in these proceedings).

These advantages of photon interferometry have stimulated
theoretical and experimental studies, despite the technical
difficulties due to the small rates of photon production and
the background due to $\pi^0$ decays. 

Besides the difference in the coupling constant, photons and
hadrons (for the sake of concreteness we shall refer in the
following to pions) have also other distinguishing
properties like spin, isospin, and mass which manifest
themselves in the corresponding BEC
and which sometimes are overlooked.
 This is the subject of this talk .
 
\section{Comparison between photon and hadron BEC}       

\begin{table} 
\begin{center}
{\large{\bf Table 1}}\\ [1.0cm]
{\large{\bf Photons versus hadrons}} \\ [0.5cm]

\begin{tabular}{lll}
%\hline
   
\bf Photons  & \bf Properties & \bf Hadrons(Pions)\\
[0.5cm]

Trivial & Classical fields & Remarkable\\
(electromagnetic) & &(Higgs,sigma meson) \\[0.5cm]
Remarkable & Quantum fields & Trivial\\[0.5cm]
Trivial& Chaos & Remarkable\\[0.5cm]
Lasers & Condensates & Pion condensates\\[0.5cm]

No  & Final state  & Yes\\
& interactions\\[0.5cm]
$m=0$ & Mass & $m \not= 0$ \\[0.5cm]  
Yes, if effective  & Multiparticle  & Yes, if energy\\
coupling is big  & production & is big enough\\
enough (lasers)\\[0.5cm]
$S=1$ & Spin & $S=0$\\[0.5cm]
$I=0$ & Isospin & $I=1$\\[1.5cm]
$1/3 \leq C_2 \leq 3$ & {\bf Correlations} & $2/3 \leq C_2
\leq 2, $  
(for charged pions)\\ 
 &  & $1/3 \leq C_2 \leq 3 $ (for neutral pions)\\[0.5cm]
Astronomy,  &  {\bf Applications} & Particle and\\
gravitational waves,& & nuclear physics,\\
 quantum optics,  & & search for\\
 atomic physics,& & quark-gluon plasma,\\ 
chemistry, & & determination of \\
 biochemistry &  & the mass of the W \\[0.5cm]

%\hline
\end{tabular}
\end{center}
\end{table}

Table 1 contains an enumeration of differences
between photons and pions which appear relevant
from the point of view of intensity interferometry. 
We will comment upon three topics in the following  
\footnote{For more details cf. e.g. a forthcoming textbook
on Bose-Einstein correlations by the author to be published
by J. Wiley and Sons in 1999.}: classical versus quantum 
fields, condensates, and the role of spin in photon BEC.

\subsection{Classical versus quantum fields; coherence and
chaos}
As is well known BEC are sensitive to the amount of coherence
of the source and this makes intensity
interferometry a useful tool in the determination of
coherence, both for photons and for hadrons. While classical
fields are always coherent, quantum fields may be coherent
or chaotic.      
Electromagnetic
fields which are at the basis of optical phenomena 
 are ``classical", i.e.
quantum phenomena do not play a role there (the Planck
constant $h$ does not appear in the Maxwell equations).
Therefore the discovery of photons i.e. of quanta of light
was so important, as it lead to the creation of modern
quantum physics. Particle physics developped
much later and
it was
 quantum from the very beginning. Therefore the
fact that the associated particle fields are quantized is
from this point of view trivial. On the other hand in the
seventies 
it became clear that the symmetries observed in particle
physics are spontaneously broken. This fact,
which was brilliantly confirmed by the discovery of intermediate
bosons, lead via the
Golsdstone-Higgs-Kibble mechanism  necessarily to 
classical fields. Hence in particle
physics the existence of classical fields is far from trivial.
With chaos the situation is rather inversed. Conventional
optical sources are thermal and therefore chaotic. However 
in particle physics where the wave lengths of particles are
of the order of the dimensions of the sources and the
lifetime of sources may be small compared with the time
necessary for equilibration, 
  one would
expect coherence as a rule and thermal equilibrium as
exceptional. 
  
\subsection{Condensates} One of the most important 
effects of quantum optics which is
based on 
coherence is the phenomenon of {\em lasing}. Lasers are
 Bose condensates and it has been
speculated that such condensates, in particular pion
condensates, may exist also in nuclei
(cf. e.g. \cite{Migdal}) or be created in heavy ion reactions
(cf. e.g. \cite{Pratt1}, \cite{Ornik}). 

 However there exist important 
differences between photon condensates i.e. lasers and pion
condensates. Furthermore there are different theoretical
approaches to the problem of pion condensates and 
some confusing statements
 as to how pion condensates are produced. 
In the following we shall discuss briefly
 these issues. 

\subsubsection{Lasers versus pion condensates; pasers?}

BEC for inclusive processes, which constitute by far the
most interesting and most studied reactions both 
with hadrons and photons have to be treated by quantum field
theory, which is the appropiate formalism when the number of
particles is not conserved. For certain purposes however, sometimes one is interested in considering events with a fixed
number of particles. Thus the number of particles in a given event can help
selecting central collisons with small impact parameter.
Theoretically this situation can be handled within field
theory, using the methods of quantum statistics \cite{Shih}.
On the other hand for the construction of event generators wave
functions appear so far to be a convenient tool and
therefore, and also for historical reasons,
 some
theorists have continued to use the ``traditional" method of
wave function (wf), as introduced in the original GGLP paper. This implies
the explicit symmetrization of the products of single
particle wf, while in field theory the
symmetrization (of amplitudes) is authomatically achieved through the
commutation relations of the field operators. When the
multiplicities are large, the symmetrization of the wf
 becomes tedious.
This lead  
 Zajc \cite{Zajc}  
 to use numerical Monte
Carlo techniques for estimating $n$ particle symmetrized 
probabibilities, which he then applied to calculate
  two-particle BEC. He was thus able also to study the
question of the dependence of BEC parameters 
on the multiplicity $n$.
Using as input a second order BEC function parametrized in the form
%(\ref{eq:simple})
\be
C_2 \sim 1+ \lambda \exp(-{\bf q}^{2} {\bf R}^2),  
\label{eq:C2}
\ee
where ${\bf q}$ is the momentum transfer and ${\bf R}$ the
radius, Zajc found, 
and this was confirmed in \cite{Shih}, that the
``incoherence" parameter
$\lambda$ decreased with increasing $n$ \footnote{In
\cite{Zajc} the clumping in phase space due to Bose symmetry
was also illustrated;}.
   
However Zajc did not consider that this
effect means that events with higher pion
multiplicities are denser and more coherent. On the
contrary he warned against such an 
interpretation  and concluded that his results
have to be used in order to eliminate the {\em bias} introduced by
this effect into experimental observations. 
\footnote{The same
interpretation of the multiplicity dependence of BEC was
given in \cite{Shih}. In this reference the nature of the 
``fake"
coherence induced by fixing the multiplicity is even
clearer, as one studies there explicitely partial coherence 
in a consistent  quantum statistical formalism.}

This warning apparently did not deter the authors of \cite{Pratt1} and 
\cite{Chao} to do just that. Ref.\cite{Pratt1} went even so far
to derive the possible existence of pionic lasers (pasers)
 from 
considerations of this type.

 Ref.\cite{Pratt1} starts by proposing an algorithm for
symmetrizing the wf which presents
the advantages that it reduces very much the computing time
when using numerical techniques, which is applicable also
for Wigner type source functions and not only plane wave 
functions,
and which for Gaussian sources provides even analytical 
results. 

Subsequently in ref.\cite{Zimanyi} wave packets were symmetrized and in
special cases the matrix density at fixed and arbitrary $n$ 
was derived in analytical form.
This algorithm was then applied to calculate the influence of
symmetrization on BEC and multiplicity distributions. As in
\cite{Zajc} it was found that 
the symmetrization produces an effective
decrease of the radius of the source, a broadening of 
the multiplicity distribution $P(n)$ and an increase of the
mean multiplicity as compared to the non-symmetrized case.
What is new in \cite{Pratt1} is (besides the algorithm)
mainly the meaning the author attributes to these results. 

In a concrete example Pratt considers
a non-relativistic source distribution $S$ in the absence of
symmetrization effects:

\be
%Pratt 8, inlocuieste p cu k
S(k,x)=\frac{1}{(2\pi R^2mT)^{3/2}}\exp\left(-\frac{k_0}{T}-
\frac{x^2}{2R^2}\right)\delta(x_0)
\label{eq:Pratt8}
\ee
where 
\ba
k_0/T=k^2/2{\Delta}^2
\ea

Here $T$ is an effective temperature, $R$ an effective
radius, $m$ the pion mass, and $\Delta$ a constant with
dimensions of momentum.

Let $\eta_0$ and $\eta $ be the number densities before 
and after symmetrization, respectively. In terms of $S(k,x)$
we have
\be
\eta_0= \int S(k,x)d^4kd^4x
\label{eq:dens}
\ee
and a corresponding expression for $\eta $ with $S$ replaced
by the source function after symmetrization.

Then one finds \cite{Pratt1} that $\eta $ increases with $\eta_0$ and
above a certain crtitical density $\eta^{crit}_0 $, 
$\eta $ diverges. This is
interpreted by Pratt as {\em pasing}.

The reader may be rightly puzzled by the fact that while $\eta $ has a
clear physical significance the number density
$\eta_0 $ and a fortiori its critical value have no physical significance, because in nature
there does not exist a system of bosons the wf of
which is not symmetrized.   
Thus contrary 
to what is alluded to in
ref.\cite{Pratt1}, this paper does not does address really the question
 how a condensate is reached.
Indeed, the physical factors which induce condensation 
are, for systems in (local) thermal and chemical equilibrium,
\footnote{For lasers the determining dynamical factor is
among other things the inversion of the occupation of atomic
levels.},
pressure and temperature and the symmetrization is contained
automatically in the form of the distribution function
\be
f=\frac{1}{\exp[(E-\mu]/T]- 1}
\label{eq:f}
\ee 
where 
$E$ is the energy and $\mu$ the chemical potential.

To realize what is going on
 it is useful to observe that the increase of 
$\eta_0 $ can be achieved by decreasing $R$ and/or $T$. Thus
$\eta_0$ can be substituted by one or both of these two physical
quantities. Then the blow-up of the number density $\eta $
can be thought of as occuring due to a decrease of $T$ and/or
$R$. However this is nothing but the well known
Bose-Einstein condensation phenomenon. 
  
While from a purely mathematical point of view the
condensation effect can be achieved also by starting
with a non-symmetrized wf and symmetrizing it 
afterwards ``by hand" , the causal i.e. physical relationship   
is different: one starts with a bosonic i.e symmetrized
system and obtains condensation by decreasing the
temperature or by increasing the density of this {\em bosonic} 
system. To obtain a pion condensate e.g., the chemical potential
has to equate the pion mass.

A scenario for such an effect in heavy ion reactions has
been proposed in \cite{Ornik}. It is based on the decay of
short lived resonances which leads to an accumulation of
pions and takes place if the hadronic (dense) matter
decouples from chemical equilibrium earlier than from
thermal equilibrium.  
In \cite{Ornik} it was found that if a pionic Bose condensate 
is formed at any stage of the collision, it can be expected 
to survive 
until pions decouple from the dense matter, and thus it can affect the 
spectra and correlations of final state pions.  
 
This effect was then studied quantitatively by solving the
equations of relativistic hydrodynamics for a fluid
which contains also a superfluid component, corresponding to
the pion condensate. From the results obtained in this way
we quote: 
in the
single inclusive transverse momentum distribution the
signature of a maximum velocity appears, which is specific
 for a superfluid system.  
The second order correlation function $C_2$ presents the typical
features of a partially coherent system
i.e. a lowering of
the intercept and a double structure, which in principle
could be quite dramatic (up to a given value of $q, C_2$
vanishes). These features are rather specific for a pion 
condensate and distinguish such a system from optical 
condensates.
\footnote{None of the ``paser" papers \cite{Pratt1}
- \cite{Zimanyi} address the crucial
issue of directional coherence which is an essential
characteristic of optical lasers. This casts doubts whether
the terminology of ``paser" is appropiate.
For a model of directional coherence, not necessarily
related to pion condensates, 
cf.\cite{FW85}; experimental hints of this effect have possibly been seen
in \cite{Zajc84}.}. 

To conclude the ``paser" topic, one must correct 
another confusing interpretation which relates to the 
 observation
made also in \cite{Zajc} that the symmetrization produces a broadening of the
multiplicity distribution (MD). In particular starting with a
Poisson MD for the non-symmetrized wf one ends up
after symmetrization with a negative binomial. While Zajc
correctly considers this as a simple consequence of Bose
statistics, ref.\cite{Pratt1} goes further and associates
this with the so called pasing effect. That such an
interpretation is incorrect is obvious from the fact that
for true lasers
the opposite effect takes place. Before ``condensing" i.e.
below threshold their
MD is in general broad and of negative binomial form corresponding to
a chaotic (thermal) distribution while above threshold the
laser condensate is produced and as such
corresponds to a coherent state and therefore is
characterized by a Poisson MD. 

\subsection{Photon interferometry. Photon spin and  
bounds of BEC.}
In this section we discuss the difference between BEC for
photons and for pions. Certain erroneous results and
statements in the recent literature will be corrected.

Following \cite{Neu} and \cite{Leo} we consider 
 a heavy ion reaction where photons are produced
through bremsstrahlung from protons in independent
proton-neutron collisions\footnote{Photon emission from
proton-proton collisions is suppressed because it is of
quadrupole form.}. The corresponding
elementary dipole currents are 

\be
%Leo 1      
j^{\lambda}(k)=\frac{ie}{mk^0}{\bf p}.{\bf
\epsilon}_{\lambda}(k)
\label{eq:Leo1}
\ee
where ${\bf p} = {\bf p}_i-{\bf p}_f$ is the difference
between the initial and the final momentum of the proton,
${\bf \epsilon}_{\lambda}$ is the vector of linear polarization
and $k$ the photon 4-momentum; $e$ and $m$ are the charge and mass
of the proton respectively. The total current is written 

\be
%Leo 2, inlocuieste y cu x
J^{\lambda}(k)=\sum^N_{n=1}e^{ikx_n}j_n^{\lambda}(k).
\label{eq:Leo2}
\ee
For simplicity we will discuss in the following only the 
case of pure chaotic currents \\$<J^{\lambda}(k)> = 0$. 
The index $n$ labels the independent nucleon collisions
which take place at different space-time points $x_n$. These
points are assumed to be randomly distributed in the
space-time volume of the source with a distribution function
$f(x)$ for each elementary collision. The current correlator
then reads

\begin{eqnarray}
%Leo 9,
<J^{\lambda_1}(k_1)J^{*\lambda_2}(k_2)>&=&
<J^{\lambda_1}(k_1)J^{\lambda_2}(-k_2)>\equiv
C^{\lambda_1\lambda_2}(k_1,k_2)\nonumber\\
&=&\sum^N_{n,m=1}\int\prod^N_{l=1}d^4x_1f(x_1)\exp(ik_1x_n-
ik_2x_m)<j_n^{\lambda_1}(k_1)j_m^{\lambda_2}(-k_2)>\nonumber\\
&=&\sum^N_{n=1}[\tilde{f}(k_1-k_2)<j_n^{\lambda_1}
(k_1)j_n^{\lambda_2}(-k_2)>-\tilde{f}(k_1)\tilde{f}(-k_2)<
j_n^{\lambda_1}(k_1)>\nonumber\\
& &<j_n^{\lambda_2}(-k_2)>]
+<J^{\lambda_1}(k_1)><J^{\lambda_2}(-k_2)>.
\label{eq:Leo9}
\end{eqnarray}

Here $\tilde{f}(k)$ is the Fourier transform of $f(x)$ with
the normalization $\tilde{f}(k=0)=1$. The function
$\tilde{f}$ has a maximum at $k=0$ and becomes usually
negligible for $kR \gg 1$ where $R$ is the effective radius
of the source.
 
We will limit further the discussion to the important case 
from the experimental point of view of unpolarized photons.
The corresponding cross
sections are obtained by summing over the the polarization
indexes and the elementary currents $j_n$. Thus the
correlator defined above will be proportional to products of
the form 
\be
%Leo10 
<J^{\lambda_1}(k_1)J^{\lambda_2}(-k_2)>={\bf
{\epsilon}}^{i}_{\lambda_1}(k_1)\left(\sum^N_{n=1}<{\bf
p}^{i}_n{\bf p}^j_n>\right){\bf {\epsilon}}^j_{\lambda_2}(k_2)
\label{eq:Leo10}
\ee

Due to the axial symmetry around the beam direction one
has for the momenta the tensor
decomposition
\be
%Leo 11
<{\bf p}^{i}_n{\bf p}^j_n>=\frac{1}{3}\sigma_n\delta^{ij}+
\delta_nl^{i}l^j,
\label{eq:Leo11}
\ee
where $l$ is the unit vector in the beam direction and
$\sigma _n, \delta _n$ are real positive constants. In
\cite{Neu} an isotropic distribution of the momenta was
assumed. This corresponds to the particular case $\delta _n = 0$. 
The generalization to the form (\ref{eq:Leo11}) is due to
\cite{Leo}. The summation over polarization indexes is
performed by using the relations

\be
<({\bf \epsilon}^{i}.{\bf p}_l)({\bf \epsilon}^{j}.
{\bf p}_{l'})>
=\frac{1}{3}({\bf \epsilon}^{i}.{\bf \epsilon}^{j})\delta_{ll'}
\label{eq:sum}
\ee

and

\be
%Leo 21
\sum^2_{\lambda=1}{\bf \epsilon}^{i}_{\lambda}(k).{\bf
\epsilon}^j_{\lambda}(k)=\delta^{ij}-{\bf n}^{i}{\bf n}^j,
\label{eq:Leo21}
\ee
where ${\bf n} = {\bf k}/ {|{\bf k}|}$.
 
We write below the results for
the second order correlation function
\be
%Leo 24
C_2(k_1,k_2)=\frac{\rho_2(k_1,k_2)}{\rho_1(k_1)\rho_1(k_2)}
\label{eq:Leo24}
\ee
 for two extreme cases:
(1) Uncorrelated elementary currents (isotropy) ($\sigma \gg
\delta$)

\be
%Leo 25
C_2(k_1,k_2;\;\sigma\neq0,\;\delta=0)=
1+\frac{1}{4}[1+({\bf n}_{1}.{\bf n}_{2})^2]\left[|\tilde{f}(k_1-k_2)|^2+
|\tilde{f}(k_1+k_2)|^2\right],
\label{eq:Leo25}
\ee
leading to an intercept 
\be
%Leo 26
C_2(k,k)=\frac{3}{2}+\frac{1}{2}|\tilde{f}(2k)|^2
\label{eq:Leo26}
\ee
limited by the values (3/2,2).  
(2) Strong anisotropy
($ \sigma  \ll \delta $):
\be
%Leo 27
C_2(k_1,k_2;\;\sigma=0,\;\delta\neq0)=
1+|\tilde{f}(k_1-k_2)|^2+|\tilde{f}(k_1+k_2)|^2
\label{eq:Leo27}
\ee
with an intercept
\be
%Leo 28
C_2(k,k)=2+|\tilde{f}(2k)|^2
\label{eq:Leo28}
\ee
limited this time by the values (2,3). 

These results are remarkable among other things because they
illustrate the specific effects of photon spin on BEC. Thus
while for (pseudo-)scalar pions the intercept is a constant
(2 for charged pions and 3 for neutral ones) even for
 unpolarized photons the intercept is a function of k.
One thus finds that, while for a system of charged pions (i.e. a mixture of $50\%$ 
positive and $50\%$ negative) the maximum value of the 
intercept Max$C_2(k,k)$ is 1.5, for
photons  Max$C_2(k,k)$ exceeds this value and this excess 
reflects the space-time properties of the source represented by
$\tilde{f}(k)$,
 the degree of (an)isotropy of the source represented by 
the quantities $\sigma$ and $\delta$, and the supplimentary
degree of freedom represented by the photon spin. The fact that 
the differences between charged pions and photons
 are enhanced for soft photons reminds us of a similar
effect found with neutral pions (cf. ref.\cite{APW}). 
Neutral pions are in general more bunched than identically 
charged ones and this difference is more pronounced for soft
pions. This similarity is not accidental, because photons 
as well as
$\pi^0$ particles are neutral and this circumstance has
 quantum field theoretical implications which will be
mentioned also below.

We see thus that in principle photon BEC can provide
information both about the space-time form of the source and
the dynamics.

These results on photon correlations refer to
the case that the sources are ``static" i.e. not expanding.
Expanding sources were considered in \cite{Feld} within a covariant
formalism. 

 The results quoted above, in particular eqs.
(\ref{eq:Leo25},\ref{eq:Leo26}),
which had been initially derived by Neuhauser, were
challenged by Slotta and Heinz \cite{Heinz}. 
Among other things, these authors claim that for photon
correlations due to a chaotic
source ``the only change relative to 2-pion interferometry
is a statistical factor $\frac{1}{2}$ for the overall
strength of the correlation which results from the
experimental averaging over the photon spin". In \cite{Heinz}
an intercept $\frac{3}{2}$ is derived which is 
in contradiction with the results presented above
and in particular with eq.(\ref{eq:Leo26}) where besides the 
factor $\frac{3}{2}$ there appears also the k dependent 
function $\frac{1}{2}{|\tilde{f}(2k)|^2}$. 

We would like to point out here that the reason for
the difference between the results of \cite{Neu},\cite{Leo}
on the one hand and those of ref.\cite{Heinz} on the other
is mainly due to the fact that in 
\cite{Heinz} a formalism was used which is less general
than that used in \cite{Neu} and \cite{Leo} and which is 
inadequate for the present problem. This implies among 
other things that unpolarized photons cannot be treated in
the naive way proposed in \cite{Heinz} and that the results
of \cite{Neu} and \cite{Leo} are correct, while the results
of \cite{Heinz} are not.
 
 In 
\cite{Heinz} 
the following formula for the second order correlation 
function is used: 
\be
%Heinz 23
C({\bf k}_1,{\bf k}_2)=1+\frac{\tilde {g}_{\mu \nu}(
{\bf q,K})\tilde {g}^{\nu \mu}({\bf -q,K})}
{\tilde {g}^{\mu}_{\mu}({\bf 0,k}_1)\tilde {g}^{\mu}_{\mu}(
{\bf 0,k}_2)}
\label{eq:heinz23}
\ee
Here $\tilde{g}$ is the Fourier transform of a source 
function $(g(x,K)$ and ${\bf q}={\bf k}_1-
{\bf k}_2$, ${\bf K}={\bf k}_1+{\bf k}_2$. 

This formula is a particular case of a more general formula
for the second order correlation function derived
 by Shuryak \cite{Shuryak}
using a model of uncorrelated sources, when 
  emission of particles from the same space-time point is
negligible. 

As is clear from this derivation
there exists also a third term, neglected in
eq.(\ref{eq:heinz23}) and which corresponds to the 
simultaneous emission of two particles from the same point
 (cf. \cite{APW}). 
While for massive
particles this term is in general suppressed, this is not
true for massless particles and in particular for soft
photons. Indeed in \cite{Neu} and \cite{Leo} this additional
term had not been neglected as it was done subsequently in \cite{Heinz}   
and therefore it is not surprising that
ref.\cite{Heinz} could not recover the results of
refs.\cite{Neu} and \cite{Leo}. The neglect of the term
corresponding to emission of two particles from the same
space-time point 
 is not permitted in the present case.
Emission of particles from the same space-time point  
corresponds in a first approximation to
particle-antiparticle correlations and this type of effect
leads also to the difference between BEC for identical 
charged pions and the BEC for neutral pions.
This is so because neutral particles coincide with the
corresponding antiparticles. (As a consequence of this
circumstance e.g.
while for charged pions the maximum of the intercept is 2, for neutral pions
it is 3 (cf. \cite{APW} and Table 1). Photons being neutral particles,
similar effects like those observed for $\pi^0$-s are
expected and indeed found.

 This misapplication of the current formalism  
 invalidates completely the conclusions of 
ref.\cite{Heinz}. 

  Intuitively the fact that for unpolarized
photons  Max$C_2(k,k)$ is 2 and not 1.5 as stated in
\cite{Heinz}, can be explained as follows: a system of 
unpolarized photons consists on the average of $50\%$
photons with the same helicities and $50\%$ photons with
opposite helicities. The first ones contribute to the
maximum intercept 
with a factor of $3$ and the last ones with a factor of 1
(coresponding to unidentical particles).  

For the sake of clarification it must be mentioned that ref.\cite{Heinz} 
contains also other incorrect 
statements. Thus   
the claim in \cite{Heinz} that the approach by Neuhauser
``does not correctly take into account the constraints from
current conservation" is completely unfounded
 as can be seen from eq.(\ref{eq:sum}) which is a
an obvious consequence of current conservation. 
Last but not least the statement that because
the tensor structure in eq.(20) of
ref.\cite{Feld} is parametrized
in terms of $k_1$ and $k_2$ separately ``instead of only in
terms of $K$, leading to spurious terms in the tensor
structure which eventually result in their spurious
momentum-dependent prefactor", has also to be qualified. 
Indeed the additional term,
unduely neglected in \cite{Heinz}, depends not only on
$K$ but also on $k_1$ and $k_2$ separately and this
contradicts the entire argumentaton of \cite{Heinz}
regarding the ``spurious terms".   

The considerations presented above refer to the effects of
photon spin on the upper bounds of the correlation function.
Similar specific effects exist also for the lower bounds
 \cite{APW} (cf. Table 1): 
$C^{--}_2(k_1,k_2)\geq 2/3$ and 
$C^{00}_2(k_1,k_2)\geq 1/3$.
Here the indeces $--$ and $00$ refer to charged and neutral
pions (photons) respectively. These lower bounds have also
lead to confusion in the literature and this issue was 
clarified and
corrected in \cite{Axel}. 
For further details of the topics discussed here cf.
\cite{RW}.

\end{document}